\documentstyle[11pt,a4,grqc]{article}
\def\Cl#1{{\cal C}\!\ell_{#1}}
\def\uls#1{{\scriptscriptstyle \underline{#1\!}\,}}
\def\ul#1{\underline#1}
\def\rCl#1{\mbox{\Large ${ }^{r}$}\!\Cl{#1}}
\def\rLambda{\mbox{\Large ${}^{r}$}\!\!\Lambda}
\def\rev{\widetilde}
\def\revg#1{(#1)\,\tilde{ }}
\def\conj#1{\overline{{#1}^{\!\raisebox{0.1cm}{\,}}} }
\def\conju#1{\,\overline{\! {#1}^{\!\raisebox{0.15cm}{\,}}} }
\newcommand{\prj}[2]{\left\langle #2 \right\rangle_{#1}}
\def\f12{\frac{1}{2}}

\def\lit#1{\mbox{\raisebox{0.03cm}{${\scriptstyle #1}$}}}
\def\*{\! *\!}
\def\ot{\mbox{\raisebox{0.03cm}{${\scriptscriptstyle\bf\otimes}$}}}
\def\no{\noindent}

\def\w{\wedge}

\def\beq{\begin{equation}}
\def\eeq{\end{equation}}
\def\inn{\!\cdot\!}

\begin{document}

\title{Clifford Algebra Approach to Superenergy Tensors}

\author{J.\ M.\ Pozo, J.\ M.\ Parra}

\address{Departament de Fisica Fonamental, Universitat de Barcelona\\
Diagonal 647, E-08028 Barcelona, Spain\\
E-mail: {\em {\tt jpozo@ffn.ub.es, jmparra@ffn.ub.es}}}

\maketitle\abstracts{
Senovilla has recently defined an algebraic construction 
of a superenergy tensor $T\{A\}$ from any arbitrary tensor $A$, by 
structuring it as an $r$-fold form. This superenergy tensor satisfies 
automatically the dominant superenergy property. We present a more 
compact definition using the $r$--direct product Clifford algebra 
$\mbox{\large ${}^{r}$}\!\Cl{p,q}$. This form for the superenergy 
tensors allows to obtain 
an easy proof of the dominant superenergy property valid for any 
dimension. }

\section{Introduction}

The name of superenergy was first applied to the Bel-Robinson
(BR) and Bel tensors defined from the conformal Weyl tensor and
the Riemann tensor respectively. The motivation for this name is
that they share many properties with energy-momentum tensors. 
The prefix super appears because they are rank-4 tensors instead 
of rank-2. 
Senovilla \cite{Seno,Seno2} has recently defined an algebraic construction of
{\it superenergy tensors} $T\{A\}$ from arbitrary {\it seed tensors} $A$,
which unifies in a single procedure the BR and Bel 
tensors and many energy-momentum tensors from different physical 
fields. 
Given any tensor $A$, its superenergy tensor $T\{A\}$ satisfies the 
{\it Dominant Superenergy Property} (DSEP). 
These superenergy tensors have been used to study the 
interchange of {\it superenergy} between different
fields and to provide a simple geometric criterion for the 
causal propagation of physical fields \cite{SenoBergq}.
Bergqvist \cite{Bergq} has found an expression with 2-spinor
formalism, valid for dimension 4, which leads to an elegant proof of the DSEP.
In this article we introduce an alternative formulation for
superenergy tensors using real Clifford algebra. Our expression is 
simpler and more compact than both, standard-tensorial and spinorial
definitions, and it allows a proof of the DSEP, as elegant as the 
2-spinors one and valid for any dimension. 

\section{Multivectors}

Let us consider the tangent vector space ${\cal T}({\cal M})$ of some real 
manifold ${\cal M}$ of dimension $n$ and signature $\{p,q\}$, with 
$n=p+q$. A multivector is an element of the Grassmann or exterior
algebra $\Lambda\equiv\Lambda({\cal T}({\cal M}))$, i.e. it is a 
linear combination of scalar, vector, bivector, etc. 
It is over this tangent multivector space $\Lambda$ that
Clifford algebra $\Cl{p,q}$ 
is built, by endowing $\Lambda$ with the Clifford geometric
product. The geometric product is an associative product which,
in contrast to the exterior product, is specific to the metric \cite{Hestenes}.

To write a multivector $A\in\Cl{p,q}$ in a complete basis of 
the linear space $\Cl{p,q}=\Lambda$ we will use a {\it multiindex}, denoted
in a latin capital letter,
\[ A=A^Ie_{I} \ \ .\]

A basic operation on multivectors is the {\it degree projection} 
$\prj{s}{A}$, which selects from the multivector $A$ its $s$-vector part 
($0$=scalar, $1$=vector, $2$=bivector \ldots ).
There are two natural antiinvolutions, which are independent of
the metric but are fixed by the graded structure of the multivectors space
$\Lambda$: {\it Reversion}, denoted 
with a tilde $\rev{A}$, and {\it Clifford con\-ju\-ga\-tion}, with an 
overline $\conju{A}$. They both reverse the order of products 
$\rev{\! AB}=\rev{B}\rev{A}$, $\conju{AB}=\conju{B}\,\conju{A}$, but
reversion keeps vectors unchanged $\rev{a}=a$, while Clifford conjugation
changes the sign of vectors $\conj{a}=-a$. Taking, for instance,
a factorizable trivector
\[ \revg{a\w b\w c}=c\w b\w a=-a\w b\w c \ ,\hspace{0.7cm}
\conj{a\w b\w c}=(-c)\w(-b)\w(-a)=a\w b\w c \]

\section{${\bf r}$-fold multivectors}

An {\it $r$-fold multivector} A is an element of the 
$r$-direct product of the Clifford algebra $\Cl{p,q}$.
\[ 
A\in \rCl{p,q}\equiv 
\underbrace{\Cl{p,q}\ot\Cl{p,q}\ot\cdots\ot\Cl{p,q}}_r =
\underbrace{\Lambda\ot\cdots\ot\Lambda}_r\equiv \rLambda
\]

\no Each multivector space $\Lambda$ will be called a {\it
block}. Its expression in a basis, using multiindices, is
\beq \label{r-fold}
A=A^{I_1 \, I_2 \ \cdots \ I_r} \,
e_{I_1}\ot\, e_{I_2}\ot\,\cdots\,\ot\, e_{I_r} \ \ .
\eeq

The natural associative product defined between two $r$-fold 
multivectors is the {\it $r$-direct Clifford product}, which involves 
an independent Clifford product in each block.
\[
A B=\left( A^{I_1 \,\cdots\, I_r} \,e_{I_1}\ot
\cdots\ot\, e_{I_r}\right)\,\left(B^{J_1 \,\cdots\, J_r}
\, e_{J_1}\ot\cdots\ot\, e_{J_r}\right) 
\]
\[
=A^{I_1 \,\cdots\, I_r}B^{J_1 \,\cdots\, J_r}\,(e_{I_1}e_{J_1})
\,\ot\,\cdots\,\ot\, (e_{I_r}e_{J_r})
\]

To shorten expressions we introduce {\it multi-fold multiindices}.
These collect a list of multiindices and are denoted with an 
underline, $ \ul{I}\equiv\{I_1,I_2,\cdots ,I_s\} $.
Using them the expression (\ref{r-fold}) simplifies to
$ A=A^{\,\uls{I}} \, e_{\uls{I}} $.
Multi-fold multiindices with $s<r$ will be used to make a
single block explicit.
\[
A=A^{\uls{C}\, I_i\, \uls{D}} \, 
e_{\uls{C}}\,\ot\, e_{I_i}\ot\, e_{\uls{D}} =
e_{\uls{C}}\,\ot \;\! A^{\uls{C}\,\uls{D}}\,\ot\, e_{\uls{D}}
\]

\no where $\ul{C}=\{I_1,\cdots ,I_{i-1}\}$ and  
$\ul{D}=\{I_{i+1},\cdots ,I_r\}$. Note that 
$A^{\uls{C}\,\uls{D}}\equiv A^{\uls{C}\, I_i\, \uls{D}}\, e_{I_i}$
is not just a scalar component but a multivector.

The basic operations of degree projection, reversion and Clifford 
conjugation, acting on multivectors, can be extended to $r$-fold
multivectors. Thus, we define the {\it $r$-fold degree projection},
the {\it $r$-fold reversion} and the {\it $r$-fold Clifford 
conjugation} as the result of applying the corresponding
operation independently on every block, and will use the same
notation as with simple multivectors. 
\[ 
\prj s{A}=A^{I_1 \, I_2 \, \cdots \, I_r} \prj s{e_{I_1}}\ot
\prj s{e_{I_2}}\ot\,\cdots\,\ot \prj s{e_{I_r}} \ \ ,
\]
\[
\rev{A}=A^{I_1 \, I_2 \,\cdots\, I_r}\,\rev{e_{I_1}}\ot
\,\rev{e_{I_2}}\ot\,\cdots\,\ot\,\rev{e_{I_r}}
\mbox{ \ \ \ and \ \ \ } 
\conju{A}=A^{I_1 \, I_2 \,\cdots\, I_r}\,\conj{e_{I_1}}\ot
\,\conj{e_{I_2}}\ot\,\cdots\,\ot\,\conj{e_{I_r}} 
\]

\section{Superenergy tensors}

Given an arbitrary tensor
\[ 
\hat{A}=\hat{A}^{\mu_1 \,\cdots\, \mu_s}\,e_{\mu_1}
\ot\,\cdots\,\ot\, e_{\mu_s},
\]

\no the procedure defined by Senovilla gives an associated
{\it basic superenergy tensor} $T\{A\}$.
The seed tensor $\hat{A}$ is treated as an 
{\it $r$-fold form}, by reordering and grouping 
antisymmetric indices in separated blocks.
Thus, the {\it reordered tensor} $A$ is an
{\it $r$-fold ($n_1,n_2,\ldots,n_r$)-form}.
\[
A\in\ \Lambda^{n_1}\ot\Lambda^{n_2}\ot\cdots\ot\Lambda^{n_r}
\ \subset \ \ 
\rLambda=\rCl{p,q}
\]

\no We say that $A$ is a {\it degree defined} $r$-fold multivector.

An energy condition has sense only for Lorentzian 
manifolds, that is, for signatures 
$\{1,p\}$ or $\{p,1\}$.
Here we will use the signature $\{p,1\}$, although everything
can be done in the reversal signature without any trouble. 

Senovilla's definition for the basic superenergy 
tensor has the following form \cite{Seno}
\beq 
\label{Tens_def}
T\{A\}=\f12 \sum_{\cal P} A_{\cal P}\times A_{\cal P} 
\eeq

\no where $A_{\cal P}$ denote the $r$-fold form $A$
transformed by a combination of duals: that is, ${\cal P}$ codifies
taking the Hodge dual on some blocks and keeping the rest of blocks
unchanged. Thus, the summation runs through the $2^r$ possible 
combinations. The cross product $A\times A$ is defined here as 
the contraction in every block of all the indices except one in each 
factor. Therefore, $T\{A\}$ has $r$ pairs of indices:
\beq
\label{cross}
(\!A\!\times\! A)_{\mu_1 \nu_1 \ \cdots \ \mu_r \nu_r}\!=\! 
\left(\,\prod_{i=1}^r
\frac{1}{(n_i-1)!}\right)
A_{\mu_1 \,\lambda_{12}\cdots \lambda_{1n_1}\ \cdots \  
\mu_r \,\lambda_{r2}\cdots \lambda_{rn_r}}
\ A_{\nu_1}{}^{\lambda_{12}\cdots \lambda_{1n_1}}
{}^{\ \cdots}_{\ \cdots \ \nu_r}{}^{\lambda_{r2}\cdots \lambda_{rn_r}}
\eeq

The use of the $r$-fold Clifford algebra $\rCl{p,1}$ allows us
to introduce an alternative and much more compact definition of
the basic superenergy tensor.  Our expression, inspired by the
Clifford geometric algebra formulation of the electromagnetic
stress-energy \cite{Hestenes}, defines the tensor
directly applied to $r$ pairs of vectors.
\beq
T\{A\}(u_1\ot u_2\ot\cdots\ot u_r)(v_1\ot v_2\ot\cdots\ot v_r)
\ = \ 
(-1)^r\f12\prj0{A \ (u_1\ot\cdots\ot u_r) \,\conju{A}\ (v_1\ot\cdots\ot v_r)}
\label{Clif_def}
\eeq 

\no If we are interested in its components in a basis 
$\{e_\mu\}$, we simply apply the tensor to the corresponding basis elements.
\beq
\label{Components}
{T\{A\}}_{\mu_1\,\nu_1 \ \cdots \ \mu_r\,\nu_r}=(-1)^r\f12
\prj0{A \ (e_{\mu_1}\ot\cdots\ot\, e_{\mu_r})\,\conju{A}\ 
(e_{\nu_1}\ot\cdots\ot\, e_{\nu_r})} 
\eeq

In the remaining part of the section we will show that both
expressions correpond to the same object. To this purpose let us
start with the Clifford algebra 
definition (\ref{Clif_def}) and expand it to obtain the standard-tensorial 
definition (\ref{Tens_def}). Let us concentrate on a single arbitrary block. 
Using multi-fold multiindices, the components (\ref{Components})
can be written as 
\[ 
T\{A\}_{\mu_1 \nu_1 \cdots \mu_r \nu_r}= 
\]
\[ 
(\!-1\!)^r\f12\!
\prj0{e_{\uls{C}}e_{(\mu_1,\cdots,\mu_{i-1})}
\conj{e_{\uls{E}}}e_{(\nu_1,\cdots,\nu_{i-1})}}
\prj0{A^{\uls{C}\uls{D}}\ e_{\mu_i}\ 
\conju{A^{\uls{E}\uls{F}}}\ e_{\nu_i}}
\prj0{e_{\uls{D}}e_{(\mu_{i+1},\cdots,\mu_r)}
\conj{e_{\uls{F}}}e_{(\nu_{i+1},\cdots,\nu_r)}} 
\]

To reexpress the result in that $i$th block we take into account
two essential facts. The first
is that the Clifford product of any multivector with a vector can be 
splitted into inner and exterior products.
\[
A^{\uls{C} \uls{D}}\ e_{\mu_i} = 
A^{\uls{C} \uls{D}}\cdot e_{\mu_i} +  
A^{\uls{C} \uls{D}}\w e_{\mu_i} 
\]
With this expansion the block
$\langle {(A^{\uls{C}\uls{D}}\, e_{\mu_i})\ 
(\conju{A^{\uls{E}\uls{F}}}\, e_{\nu_i})} 
\rangle_{{}_0}$
splits also into $2$ terms, since the cross terms vanish.
\[
\prj0{A^{\uls{C} \uls{D}}\ e_{\mu_i}\ 
\conju{A^{\uls{E} \uls{F}}}\ e_{\nu_i}}=
\prj0{ ( A^{\uls{C} \uls{D}}\cdot e_{\mu_i})\,
( \conju{A^{\uls{E} \uls{F}}}\cdot e_{\nu_i}) }
+ \prj0{ ( A^{\uls{C} \uls{D}}\w e_{\mu_i})\,
( \conju{A^{\uls{E} \uls{F}}}\w e_{\nu_i}) }
\]

The second fact is that an exterior product can be written, with
the help of Hodge duality, as an inner 
product. Applying it, the second term has the same structure as
the first term, but where in the first 
we have the original $n_i$-form, $A^{\uls{C} \uls{D}}$, in the second
we have the dual $(n\!-\!n_i)$-form, $*A^{\uls{C} \uls{D}}$.
\[
\prj0{A^{\uls{C} \uls{D}}\ e_{\mu_i}\ 
\conju{A^{\uls{E} \uls{F}}}\ e_{\nu_i}}=
\prj0{ \big( A^{\uls{C} \uls{D}}\cdot e_{\mu_i}\big)\,
\big( \conju{A^{\uls{E} \uls{F}}}\cdot e_{\nu_i}\big) } + 
\prj0{\big( [*A^{\uls{C} \uls{D}}]\cdot e_{\mu_i}\big)\,
\big( \conju{[*A^{\uls{E} \uls{F}}]\!}\cdot e_{\nu_i}\big) }
\]

Repeating this expansion for every block we obtain $2^r$
terms, corresponding to all possible combinations that take 
the dual in some blocks and leave unchanged the rest.
\beq
\label{Summation}
T\{A\}_{\mu_1 \nu_1 \cdots \mu_r \nu_r}=
\f12\sum_{\cal P}(-1)^r
\prj0{ ( A_{\cal P}\cdot \lit(e_{\mu_1}\ot\cdots\ot\, 
e_{\mu_r}\lit) ) \ ( \conju{A_{\cal P}}\cdot 
\lit(e_{\nu_1}\ot\cdots\ot\, e_{\nu_r}\lit)) }
\eeq

\no where the dot denotes the inner product in every block, and we have
used ${\cal P}$ again to indicate each combination of Hodge duals.
Finally, comparing this last expression with (\ref{Tens_def}), we only have
to check that the terms of this summation (\ref{Summation}) coincide with the
Senovilla's cross product (\ref{cross}).
\[
(-1)^r\prj0{(A\cdot e_{\mu_1,\ldots,\mu_r})\,
(\conju{A}\cdot e_{\nu_1,\ldots,\nu_r})}
\ = \ (A\times A\,)_{\mu_1\nu_1\ldots\mu_r\nu_r} 
\]

\no This can be seen by realizing that the dot products in (\ref{Summation})
fix one index in each factor for each block. The scalar
projection of the product selects the terms that correspond to
the contraction of the rest of the indices. It is easy to 
check that the signs coincide. Then, the proof is complete.
\[
(-1)^r\prj0{(A\inn e_{\ul{\mu}})
(\conju{A}\inn e_{\ul{\nu}})}=
\prj0{(A\inn e_{\ul{\mu}})
(e_{\ul{\nu}}\inn \rev{A})}=
\prj0{(A\inn e_{\ul{\mu}})
\rev{(A\inn e_{\ul{\nu}})}}
\]

\section{Dominant superenergy property (DSEP)}

In this section we present a simple proof of the DSEP for the 
superenergy tensor $T\{A\}$, using its expression in the $r$-fold 
Clifford algebra $\rCl{p,1}$. A superenergy tensor satisfies the 
DSEP if for all collection $\{u_i,v_i\}$ of causal and
future-pointing (f-p) vectors 
\[
T\{A\}(u_1\ot\cdots\ot u_r)(v_1\ot\cdots\ot v_r)=
(-1)^r\f12\prj0{A \ (u_1\ot\cdots\ot u_r) \,\conju{A} \ 
(v_1\ot\cdots\ot v_r)} \geq 0 \ \ .
\]

Let us recall, first, that a time-like f-p vector $u$ can always 
be expressed as the result of applying a local Lorentz transformation 
and a dilation to a chosen unitary time-like f-p vector $e_0$.
This transformation is performed by means of an even multivector inside the
same Clifford algebra.
\[ u=R_u e_0 \conju{R_u} \ \ ,\hspace{1cm} R_u\in\Cl{p,1}^+ \]

\no The same expression applies for a null vector $u$ with $R_u$ a
singular transformation. For the tensor product of $r$ f-p vectors
$\ul{u}\equiv u_1\ot u_2 \ot\cdots\ot u_r\in\rCl{p,1}$, 
the operator which implements the transformation also belongs to $\rCl{p,1}$
\[
R_{\ul{u}}\equiv R_{u_1}\ot R_{u_2}\ot\cdots\ot R_{u_r} \ 
\in\rCl{p,1}\ , \hspace{0.5cm} 
\ul{e_0}\equiv\underbrace{e_0\ot\cdots\ot e_0}_r\ ,\ \ {\rm so} 
\hspace{0.4cm}
\ul{u}=R_{\ul{u}} \ \ul{e_0} \,\conju{R_{\ul{u}}}
\]

Our proof of the DSEP proceeds in two steps. First, using the operators 
$R_{\ul{u}}\in\rCl{p,1}$, we express
the result of applying $T\{A\}$ to any set of $2r$ f-p vectors 
$\ul{u},\ul{v}$, as the {\small $\{0,\ldots,0\}$} component of another 
superenergy tensor $T\{A'\}$.

\begin{eqnarray*}
2 T\{A\}(\ul{u})(\ul{v})&=&(-1)^r
\prj0{A \ \ul{u}\,\conju{A} \ \ul{v} \,}\ \ 
=\ \ (-1)^r
\prj0{A\, \Big(R_{\ul{u}} \ \ul{e_0} \,\conju{R_{\ul{u}}}\Big) \,
\conju{A}\,\Big(R_{\ul{v}} \ \ul{e_0} \,\conju{R_{\ul{v}}}\Big)\, }\\
&=&(-1)^r
\prj0{\Big( \conju{R_{\ul{v}}} A R_{\ul{u}}\Big) \ \ul{e_0} \ 
\conju{ \:\!\Big( \conju{R_{\ul{v}}} A R_{\ul{u}} \Big) \!}\,\ 
\ul{e_0} } \ \ = \ \ \ 2 T\{A'\}(\ul{e}_0)(\ul{e}_0)
\end{eqnarray*}

\no where $A'\equiv\conju{R_{\ul{v}}} A R_{\ul{u}} \in\rCl{p,1}$ 
is also an $r$-fold multivector. 

The second step proves that this component is non negative 
for any $A'\in\rCl{p,1}$.
\[
T\{A'\}(\ul{e}_0)(\ul{e}_0)=
\f12 (-1)^r\prj0{A'\ \ul{e_0}\,\conju{A'}\ \ul{e_0}}
\geq 0 \ \ \ \ \forall A'\in\rCl{p,1}
\]

\no To see this we realize a splitting of $A'$ into parts orthogonal 
and parallel to the direction $e_0$. This splitting corresponds to the 
isomorphism of linear spaces, though not as algebras,
\[
\rCl{p,1}\simeq\rCl{p,0}\otimes\rCl{0,1}
\]

\no where $\rCl{0,1}$ is the space generated by vector $e_0$ and 
$\rCl{p,0}$ is the space generated by the Euclidean space orthogonal 
to $e_0$. A basis for $\rCl{0,1}$ has the $2^r$ elements:
\[ 
\{e_{\ul{P}}\}=\{\ 1\ot\cdots\ot 1\ot 1\ ,
\ 1\ot\cdots\ot 1\ot\;\! e_0\ ,\ 1\ot\cdots\ot\;\! e_0\ot 1\ ,
\ \ \ldots\ \ ,\ e_0\ot\cdots\ot\;\! e_0\ot\;\! e_0\  \} 
\]

\no Now, expanding $A'$ in the basis $\{e_{\ul{P}}\}$ of $\rCl{0,1}$
with components in $\rCl{p,0}$
\[
A'=A'^{\ul{P}}\ e_{\ul{P}}\ \ ,
\hspace{1cm}A'^{\ul{P}}\in\rCl{p,0}
\ \ , \hspace{1cm} e_{\ul{P}}\in\rCl{0,1}
\]

\no we finally complete the proof
\[
(-1)^r\prj0{A'\ \ul{e_0}\,\conju{A'}\ \ul{e_0}}
=\prj0{\Big(A'^{\ul{P}}\ e_{\ul{P}}\,\Big)\,\ul{e_0}\, 
\Big(\conj{e_{\ul{Q}}}\,\conju{{A'}^{\ul{Q}}}\,\Big)\, {\ul{e_0}}^{-1}}
= \prj0{\Big({A'}^{\ul{P}}\ e_{\ul{P}}\,\Big)\, 
\Big(\conj{e_{\ul{Q}}}\,\rev{{A'}^{\ul{Q}}}\,\Big)}
\]
\[
=\sum_{\ul{P}}\prj0{{A'}^{\ul{P}}\ \rev{{A'}^{\ul{P}}}\,}\geq 0
\]

\no The summation is always positive since, $\forall B\in\rCl{p,0}$, 
which is an algebra generated by an Euclidian metric, $\prj0{B\rev{B}\,}$ is 
a positive defined norm.

\vspace*{-2pt}
\section*{Acknowledgments}
This work has been supported by the scholarship AP96-52209390
and the project PB96-0384 from the Spanish Ministry of Education
and from the Catalan Physical Society (IEC).

\end{document}